\documentclass[fleqn,usenatbib]{mnras}

\usepackage{mathptmx}
\usepackage[T1]{fontenc}
\usepackage{ae,aecompl}
\usepackage{booktabs, topcapt}

\usepackage{graphicx}	% Including figure files
\usepackage{amsmath}	% Advanced maths commands
\usepackage{amssymb}	% Extra maths symbols
\usepackage{verbatim}
\usepackage[scientific-notation=true]{siunitx}
\title[Odd Harmonics in Exoplanet Photometry]{Odd Harmonics in Exoplanet Photometry:\\ Weather or Artifact?}
\author[Cowan et al.]{
Nicolas B.\ Cowan$^{1,2}$\thanks{Email: nicolas.cowan@mcgill.ca}
Victoria Chayes,$^{3}$
\'Elie Bouffard,$^{4,2}$
Max Meynig,$^{3}$
Hal M. Haggard$^{3}$\\
\\
$^{1}$Department of Earth \& Planetary Sciences, McGill University, 3450 rue University, Montreal, QC, H3A 0E8, CAN and \\ Department of Physics, McGill University, 3600 rue University, Montreal, QC, H3A 2T8, CAN.\\
McGill Space Institute, 3550 rue University, Montreal, QC, H3A 2A7, CAN.\\
$^{2}$Institut de recherche sur les exoplan\`etes, Universit\'e de Montr\'eal, C.P. 6128, Succ.\ Centre-ville, Montr\'eal QC H3C 3J7, CAN\\
$^{3}$Bard College, 30 Campus Road, Annondale-On-Hudson, NY 12504, USA\\
$^{4}$D\'epartement de physique, Universit\'e de Sherbrooke, Sherbrooke, Qu\'ebec, J1K 2R1, CAN
}

\date{Accepted XXX. Received YYY; in original form ZZZ}

\pubyear{2016}

\begin{document}
\label{firstpage}
\pagerange{\pageref{firstpage}--\pageref{lastpage}}
\maketitle

\begin{abstract}
Photometry of short-period planetary systems allows astronomers to monitor exoplanets, their host stars, and their mutual interactions. In addition to the transits of a planet in front of its star and the eclipses of the planet by its star, researchers have reported flux variations at the orbital frequency and its harmonics: planetary reflection and/or emission and Doppler beaming of starlight produce one peak per orbit, while ellipsoidal variations of a tidally distorted star and/or planet produce two maxima per orbit. Researchers have also reported significant photometric variability at three times the orbital frequency, sometimes much greater than the predictions of tidal theory. The reflected phase variations of a homogeneous planet contains power at even orbital harmonics---important for studies of ellipsoidal variations---but cannot contain odd orbital harmonics. We show that odd harmonics can, however, be produced by an edge-on planet with a time-variable map, or an inclined planet with a North-South (N-S) asymmetric map. Either of these scenarios entail weather: short-period planets are expected to have zero obliquity and hence N-S symmetric stellar forcing.  North-South asymmetry in a giant planet would therefore suggest stochastic localized features, such as weather.  However, we find that previous claims of large-amplitude odd modes in \emph{Kepler} photometry are artifacts of removing planetary transits rather than modeling them. The only reliable claims of odd harmonics remain HAT-P-7b and Kepler-13Ab, for which the third mode amplitude is 6--8\% of the planetary flux.  Although time-variable albedo maps could in principle explain these odd harmonics, upper-limits on the infrared variability of hot Jupiters make this hypothesis unlikely. We urge theorists to study the effects of close-in planets on stellar atmospheres, as this remains the only plausible hypothesis.    
%note: needs to be under 250 words
\end{abstract}


%%%%%%%%%%%%%%%%%%%%%%%%%%%%%%%%%%%%%%%%%%%%%%%%%%

%%%%%%%%%%%%%%%%% BODY OF PAPER %%%%%%%%%%%%%%%%%%

\section{Introduction}

The time-varying brightness of a planetary system reveals myriad details about the star and planets therein.  Transits of a planet in front of its host star appear as regular dips in flux: the period of the transits corresponds to that of the planet and their depth is the cross-sectional size of the planet with respect to its host star; eclipses of the planet behind its host star appear as shallower dips, the depth of which indicate the brightness of the planet compared to that of its star.  

In addition to these relatively brief events, the time-resolved photometry of a planetary system may also exhibit periodic, approximately sinusoidal, oscillations predominantly at the orbital frequency, $n=1$, and twice this frequency, $n=2$: phase variations due to reflected starlight and/or planetary emission produce variations at the orbital frequency of the planet with maximum flux near superior conjunction, while Doppler beaming of the starlight enhances the flux when the star is moving towards the observer (first quarter); ellipsoidal variations of the prolate planet and star cause maxima twice per orbit, at both quarter phases.\footnote{Although rare, eccentric short period planets can exhibit more complex phase variations \citep{Langton2008, Laughlin2009, Kane2010, Lewis2010, Lewis2013, Cowan2011, Kataria2013, deWit2016}.}  

Many researchers have used the exquisite photometry of NASA's Kepler mission \citep{Borucki2016} to search for and measure the sinusoidal phase variations of both transiting and non-transiting planetary systems.  A few have reported significant power at higher-order harmonics of the orbital frequency, in particular the $n=3$ mode (three peaks and troughs per orbit). \cite{Esteves2013, Esteves2015} and \cite{Shporer2014} measured significant $n=3$ power for two of the best-characterized hot Jupiters.  In particular, \cite{Esteves2015} reported $n=3$ relative flux amplitudes of $\Delta F/F_0=1.9(2)\times 10^{-6}$ and $7.1(3)\times 10^{-6}$ in the Kepler photometry of HAT-P-7b and Kepler-13Ab, respectively ($\Delta F$ is the semi-amplitude of the third harmonic, while $F_0$ is the orbit-averaged system flux). If interpreted as solely due to the planet, these correspond to $A_3/A_0 = 0.06$ and 0.08, respectively ($A_3$ is the planet's third harmonic semi-amplitude and $A_0$ is the orbit-averaged planetary flux). 

\cite{Armstrong2015}, henceforth A\&R15, performed a uniform analysis of all Kepler Objects of Interest and reported significant $n=3$ amplitudes for 16 short-period systems. They found that the amplitude of these odd modes was too great ($10^{-6}$--$10^{-3}$) to attribute to tidal effects of the planet on the its star.  They instead hypothesized that either a) there is an error in the Kepler pipeline, b) the third harmonic contribution of tidal effects were been underestimated by orders of magnitude, or c) planetary reflection and/or emission is contributing to the third mode.

In this manuscript we critically evaluate the possibility that planets could contribute odd modes---in particular $n=3$---to the phase variations of a planetary system. In Section 2 we establish that planets with static maps and in edge-on orbits cannot produce odd modes in either reflected light or thermal emission.  In Section~3 we show that planets can produce these odd modes if i) they do not orbit perfectly edge-on, and ii) their albedo and/or temperature map is North--South asymmetric.  However, we find that the maximum amplitude one can expect from physically possible planet maps is smaller than the values reported by A\&R15. We demonstrate in Section~4 that time-variability in a map introduces $n=3$ power, even for edge-on orbits. In Section~5 we reproduce the analysis of A\&R15 and find that the large amplitudes they reported for $n=3$ are an artifact of their data analysis.  In particular, we show that removing transits and eclipses---rather than modeling them---artificially increases the amplitude of high-order harmonics by orders of magnitude. We discuss our results in Section~6, and conclude in Section~7.

\section{Odd Nullspace for Edge-On Orbits}
In this section we introduce the formalism used to analyze exoplanet brightness markings and viewing geometries from the measured total fluxes. We briefly review previous work in this setting and provide the necessary context for our main analytical result, namely that planets in edge-on orbits with static maps produce no odd Fourier harmonics beyond $n=1$; the full proof of this result appears in Appendix \ref{TheAppendix}. 

\subsection{Light Curve Formalism}
We follow \cite{Cowan2013}, henceforth CFH13, in approximating the flux from an exoplanet---thermal emission or reflected light---as the convolution of a planetary map, $M$, with a time-varying kernel, $K$:
\begin{equation} \label{convolution}
	F(t) = \oint K(\Omega, t) M(\Omega) d\Omega, 
\end{equation}    
where $\oint d\Omega$ is the integral over the surface of a sphere. The \emph{nullspace} of \eqref{convolution} refers to the set of non-zero maps, $M\ne 0$, that have no light curve signature, $F\equiv 0$.

The integral kernel $K$ is given by the visibility in the case of thermal light and the product of the visibility and illumination in the case of reflected light.  

We assume throughout this paper that the planets have zero rotational obliquity.  A zero-obliquity state is expected for short period planets based on tidal theory, is consistent with current observations of thermal emission \citep{Majeau2012}, and may eventually be tested via reflected orbital light curves \citep[e.g.,][]{Schwartz2016, Kawahara2016}.

We further assume that the planet map is \emph{static}, meaning that the albedo and/or emission map is unchanging as seen from its host star.  This probably requires that the planet have zero obliquity since otherwise it will exhibit obliquity seasons, but does not require the planet to be synchronously rotating \citep{Cowan2011, Cowan2012b, Rauscher2014}. It also precludes large-scale weather on the orbital timescale of the planet, an assumption that we revisit in Section 4.

Real spherical harmonics provide a convenient basis for decomposing the planetary map\begin{equation}\label{planetary_map}
M(\Omega) = \sum_{\ell=0}^{\infty} \sum_{m=-\ell}^{\ell} C_{\ell}^{m} Y_{\ell}^{m}(\Omega) \end{equation} and we call the resulting component light curves 
\begin{equation}\label{harmonic_lightcurve}
	F_{l}^{m}(t) = \oint K(\Omega, t) Y_{l}^{m}(\Omega) d\Omega,
\end{equation} 
the ``harmonic light curves'' after CFH13.

Since planetary light curves are periodic functions of time, it is convenient to decompose them into a Fourier series. We label the harmonics of the Fourier decomposition by an index $n$. As shown by \cite{Cowan2008} and CFH13 for thermal light curves, the index $n$ of the Fourier decomposition can loosely be identified with the index $m$ of the spherical harmonics: power at $n=2$ in the light curve implies non-zero coefficients for one or more $m=2$ maps. Moreover, two maps with different $l$ but the same $m$ produce indistinguishable light curves to within a multiplicative constant \citep[we neglect here the potential of eclipse mapping to break that degeneracy:][]{Majeau2012}. 

The $m=n$ identification is incomplete, however: all harmonic light curves with odd $l>1$ vanish, and hence some $l$ and $m$ combinations contribute no Fourier harmonics, i.e., these spherical harmonics belong to the nullspace of the convolution. In any case, it is important to distinguish $n$ and $m$ because in the case of reflected light a single $Y_l^m$ contributes to a variety of harmonics $n$ in the Fourier decomposition of the light curve. Finally, recall that $n=1$ corresponds to the fundamental mode of the light curve, while modes with $n\geq 2$ are its harmonics.

\subsection{Edge-on Light Curves}
\label{secRefLight}
\cite{Cowan2008} showed that for an edge-on viewing geometry, the thermal nullspace includes all spherical harmonics with odd $l>1$, as well as any even-$l$ and odd-$m$ combination. 

CFH13 developed analytic expressions for low-order harmonic light curves in the case of reflected light from a synchronously-rotating, spherical planet with a static map and in an edge-on orbit. We prove in Appendix~\ref{TheAppendix} that edge-on reflected light curves in this limit contain no odd modes beyond $n=1$ (the fundamental mode).  

In other words, a synchronously-rotating planet in an edge-on orbit and with a static map cannot produce odd harmonics, be it with thermal emission or reflected light.  

\section{Odd Modes from N--S Asymmetric Planets on Inclined Orbits}
The nullspace is more limited for inclined orbits: these planets \emph{can} produce odd harmonics, provided that they have North-South asymmetric emission/albedo maps.  We use the ratio of the third mode to the orbit-averaged planetary flux, $A_3/A_0$, to quantify these odd modes. We focus on the lowest-order odd harmonic, $n=3$, because it is expected to have the greatest amplitude (Equation~\eqref{convolution} being a low-pass filter). 

\subsection{Thermal Light Curve for Inclined Planets}
As discussed extensively in CFH13, the thermal phase variations of an inclined (non-edge-on) planet may exhibit odd modes.  There is still a nullspace for odd $l>0$, but maps with even $l$ and odd $m$ are no longer in the nullspace; these maps are anti-symmetric about the equator, so an inclined viewing geometry breaks the symmetry and frees them from the nullspace.     

For thermal emission, there is a one-to-one correspondence between the $m$ of a spherical harmonic and the Fourier components of the resulting harmonic lightcurve \citep[sinusoids in longitude can be thought of as eigenvectors of the transformation;][]{Cowan2008}. In other words, $A_3$ must be due to a $Y_l^3$ component in the emission map.  Since the intrinsic brightness map of the planet must be non-negative, the lightcurve must also have a non-zero DC offset, $A_0$.  We now estimate the maximum odd amplitude one can expect compared to the mean flux of the planet: $A_3/A_0$. 

\begin{figure}
\centering
\includegraphics[scale = 0.4]{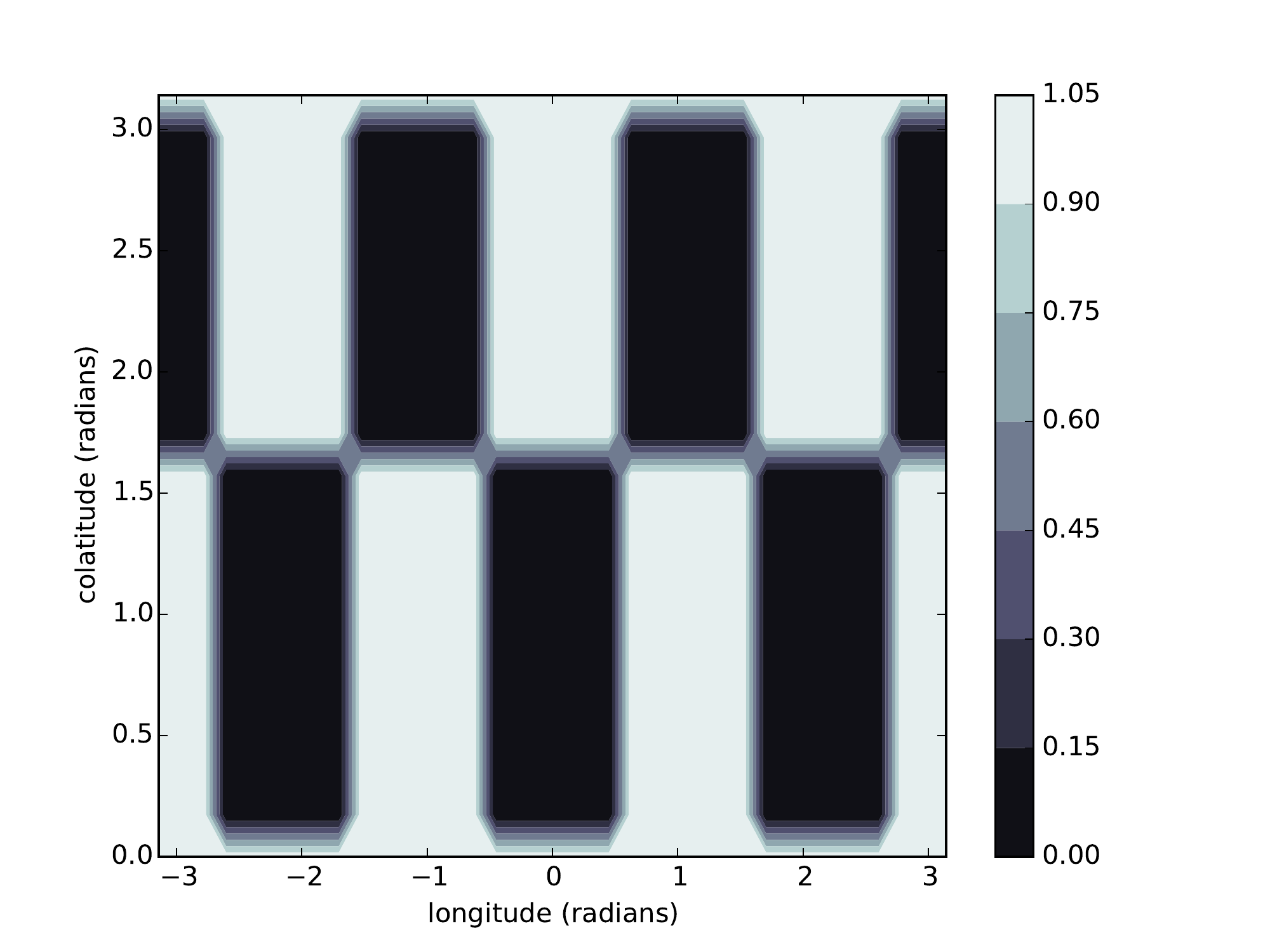}
\caption{The Boolean version of $Y_4^3$ maximizes $C_4^3$, and by extension $A_3$, for synchronously-rotating inclined planets. For thermal emission, the spherical harmonic $Y_4^3$ is the lowest-order spherical harmonic with non-zero $A_3$.  For reflected light, any spherical harmonic with odd $l-|m|$ will have non-zero $A_3$ for inclined planets, but $Y_4^3$ and $Y_4^{-3}$ contribute the most to this mode.  For the thermal emission case this should be thought of as a map of normalized emissivity, while for reflected light, this is an albedo map.}
\label{Y43_maps}
\end{figure}

The lowest-order map that contributes non-zero $A_3$ is $Y_4^3$ (CFH13); due to the low-pass nature of Equation~\eqref{convolution}, we expect (and indeed verify) that contributions to $A_3$ from $l>4$ are negligible.  We therefore construct a strictly positive map that maximizes $C_4^3$. By inspecting Equation~\eqref{convolution}, we see that the non-negative map that maximizes $C_4^3$ is a Boolean version of $Y_4^3$: where the spherical harmonic is negative, the map is zero, where the spherical harmonic is positive, the map is unity (Figure~\ref{Y43_maps}). This is analogous to maximizing a Fourier component of some frequency with a square wave of the same frequency. Since the result is intuitive, we forgo the proof using the calculus of variations.  

We then compute numerical thermal light curves for this ``Boolean $Y_4^3$'' for a variety of viewing inclinations. We use a regular grid in latitude and longitude with spatial resolution of $5^\circ$ and $15^\circ$, respectively, calculate the visibility (CFH13) over the same grid at 10,000 time steps over the course of a planetary orbit, and compute the light curve using a discretized version of Eqn 1. We assume a spherical planet, diffuse emission, and neglect limb darkening. For each light curve, we compute the Fourier coefficients and their ratios. We repeat the process for inclinations between 0 and $90^\circ$, in steps of $5^\circ$. 

\begin{figure}
\centering
\includegraphics[scale = 0.4]{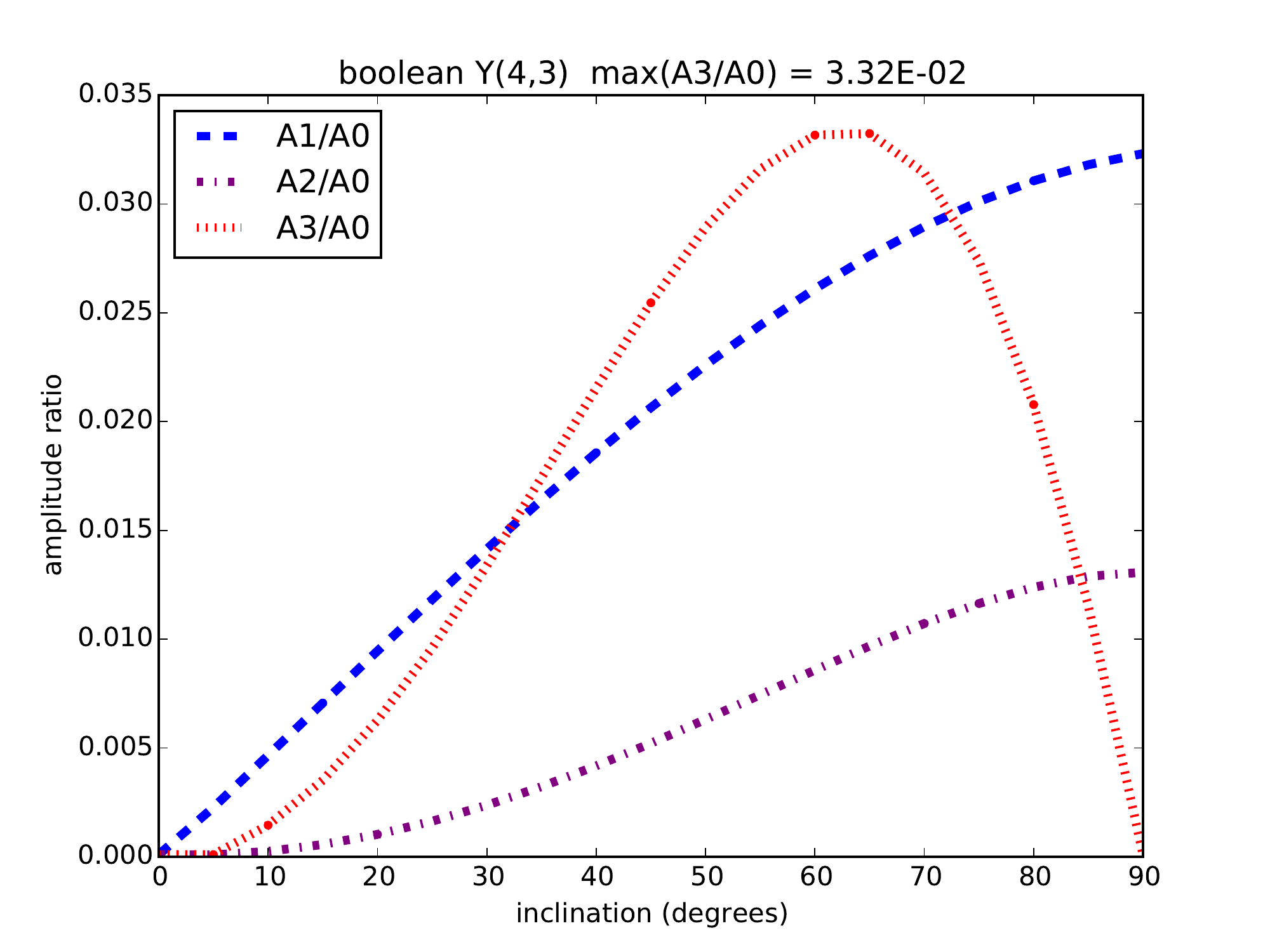}
\caption{The Fourier amplitudes of thermal light curves for the Boolean $Y_4^3$, expressed in terms of the orbit-averaged planetary flux and plotted as a function of orbital inclination. All amplitudes go to zero for pole-on viewing geometry ($i=0$) because the same planetary hemisphere remains in view. The amplitude of the fundamental mode and even harmonics increases with inclination, but the odd harmonic, $A_3$, drops back to zero for edge-on orbits ($i=90^\circ$) because the Northern and Southern hemispheres of the planet cancel out. The maximum possible $A_3$ is a few percent the orbit-averaged planetary flux, but occurs for orbital inclinations of $60^\circ$. For transiting planets, the inclination is usually $>80^\circ$, so the maximum $A_3/A_0$ is approximately 1\%.}
\label{therm_F43}
\end{figure}

In Figure~\ref{therm_F43} we show the amplitudes of Fourier components of these light curves, as a function of inclination. The maximum possible $A_3$ is a few percent the orbit-averaged planetary flux, and for transiting geometries, the maximum $A_3/A_0$ is approximately 1\%. 

\cite{Cowan2008} found that realistic limb-darkening impact mid-infrared light curves at the $<1$\% level. Limb-darkening may be more significant at the optical wavelengths relevant here, but ultimately we expect most of the visible light to be reflected starlight. We therefore deffer a quantitative study of the impacts of limb-darkening to future studies.  

\subsection{Reflected Light Curve for Inclined Planets}
\begin{figure}
\centering
\includegraphics[scale = 0.4]{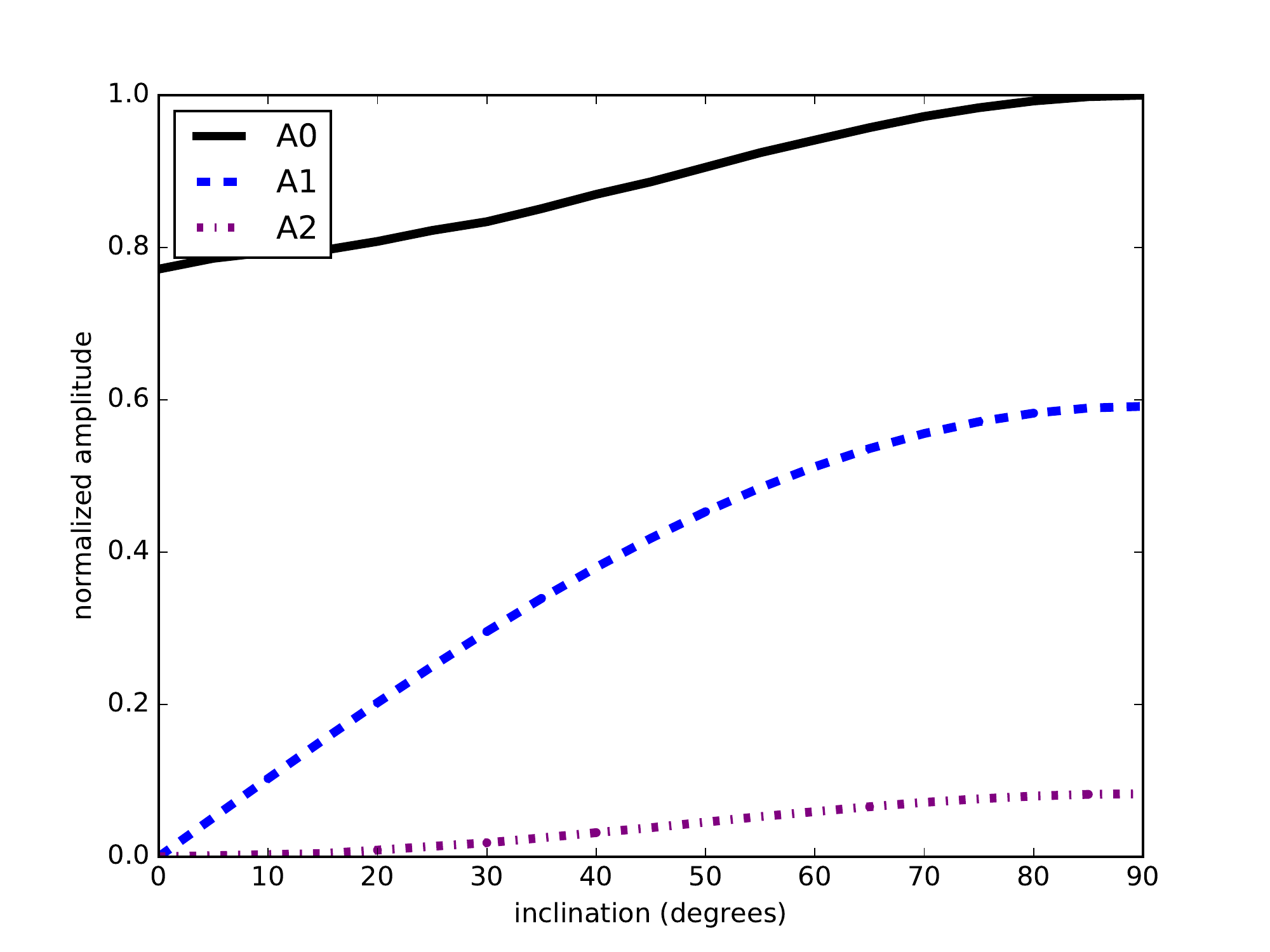}
\caption{The Fourier amplitudes of reflected light curves for a uniform-albedo, diffusely-reflecting planet as a function of orbital inclination. Although the Lambert phase curve contains no odd harmonics, it does contain significant even harmonics. The analytic Fourier decomposition of a Lambert function for edge-on orbits shows that $A_2/A_0 = 1/9 \approx 0.11$ and $A_2/A_1 = 16/(9\pi^2) \approx 0.18$. Modeling reflected light as a pure $A_1$ and ellipsoidal variations as pure $A_2$ is only accurate at the 10\% level.}
\label{Y00_Fourier}
\end{figure}

For reflected light, there is not a one-to-one correspondence between the $m$ of a spherical harmonic map and the frequency of the lightcurve. For example, a Lambert phase curve, $F_0^0$, contains significant $A_0$, $A_1$, and $A_2$, even for edge-on orbits (Figure~\ref{Y00_Fourier}). This means that a diffusely reflecting planet with uniform albedo---the bland vanilla scenario---will exhibit $n=2$ power, even in the absence of ellipsoidal variations.  Researchers who wish to use the $n=2$ amplitude as a signature of ellipsoidal variations should be cautious.  

Moreover, harmonic reflected light curves with positive and negative $m$ do not necessarily have the same Fourier spectrum.  This is again in contrast with the thermal harmonic light curves for which the negative and positive $m$ are simply related by a phase shift.

Since we do not have an analytic model for the harmonic reflected light curves of inclined synchronously rotating planets, we produce numerical harmonic light curves and compute their Fourier coefficients. We find that all N--S asymmetric spherical harmonics (odd number of meridional nodes, $l-|m|$) contribute to the third mode amplitude, $A_3$. This means that the nullspace is much more limited for reflected light than for thermal emission.    

\begin{figure}
\centering
\includegraphics[scale = 0.4]{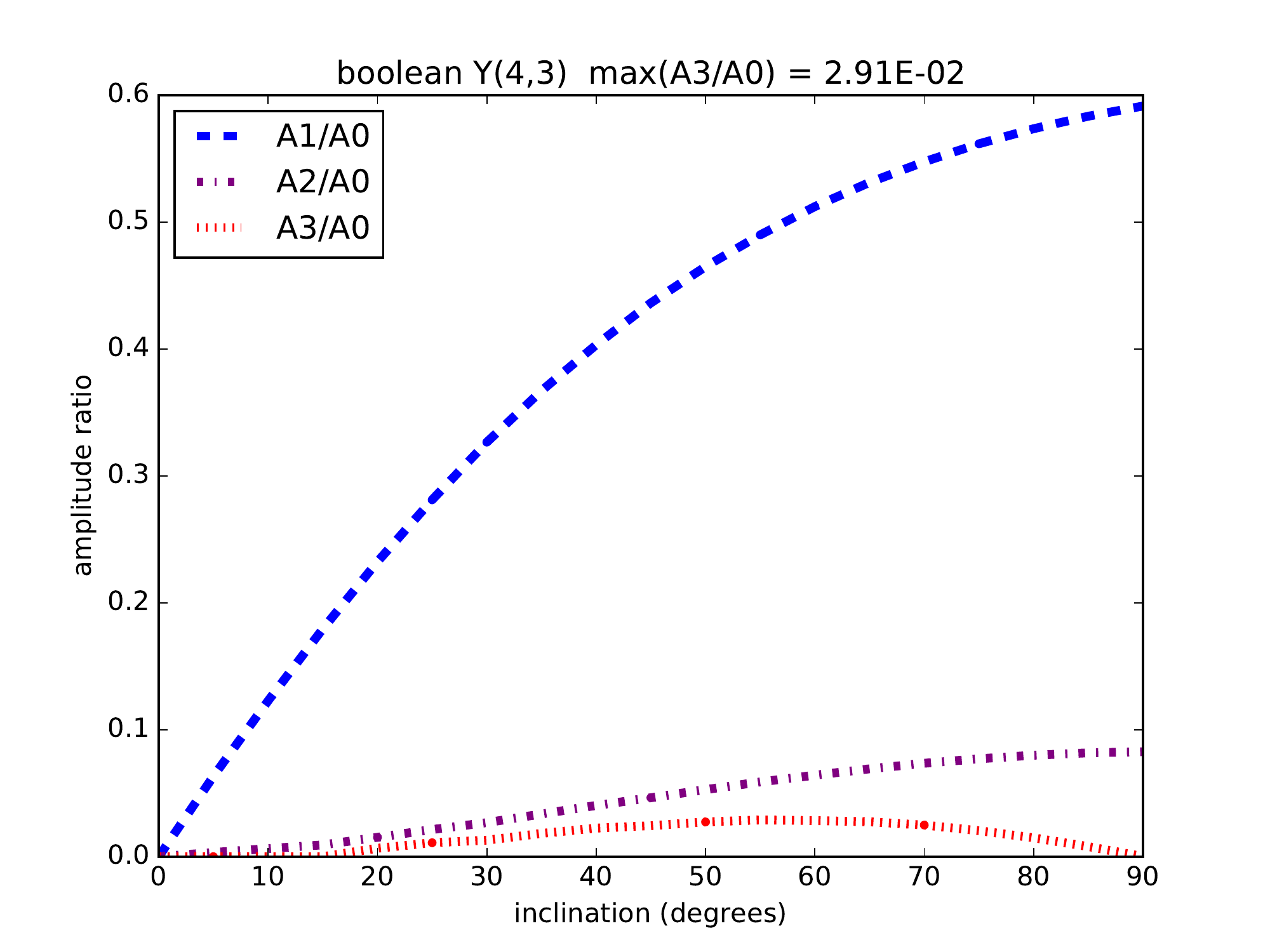}
\caption{The Fourier amplitudes of the reflected light curve for a planet with an albedo map that maximizes $Y_4^3$, plotted as a function orbital inclination (0$^\circ$ corresponds to face-on, while 90$^\circ$ corresponds to edge-on). The even harmonic is present at the $\lesssim$10\% level regardless of inclination, while the odd harmonic $A_3$ exhibits a strong inclination dependence, but is always less than 3\% the orbit-averaged planetary flux. The maximum $A_3/A_0$ one can expect for a \emph{transiting} planet (inclination $>80^\circ$) is approximately one percent.}
\label{Fourier_Y43}
\end{figure}

Although many lower-order spherical harmonics contribute non-zero $A_3$, the greatest $A_3$ again comes from $Y_4^3$, followed by $Y_4^{-3}$.  As above, we therefore consider the Boolean $Y_4^3$ (Figure~\ref{Y43_maps}). We produce synthetic reflected light curves for this map under the assumption of synchronous rotation, diffuse reflection, and for a range of orbital inclinations. We then Fourier decompose the light curves and plot the Fourier amplitudes and their ratios in Figure~\ref{Fourier_Y43}.  As with the thermal case above, $A_3/A_0$ is a few percent for inclinations of $\sim60^\circ$.  The notable difference with the thermal case is that $A_1$ is much stronger near edge-on geometries: the night side of a planet cannot reflect light, so $A_1 \lesssim A_0$.  

\section{Time-Variable Emission and Albedo Maps}
We have so far only considered static temperature and albedo maps. Here we repeat the edge-on thermal and reflected analysis of Section~2.2, but allowing the coefficients in front of the spherical harmonics to evolve linearly in time. 

The maps are still assumed to be fixed with respect to the sub-stellar point. This does not necessarily mean that the planet is synchronously-rotating: a planet may still exhibit a fixed temperature or albedo map over the course of a planetary orbit.  For example, averaged over the course of a year, Earth exhibits a relatively stable pattern with the hottest local solar time being located East of the sub-solar longitude \citep{Cowan2012b}.  For a planet with zero obliquity and eccentricity---and hence no seasons to speak of---the heating pattern is even more stable.  In other words, it is difficult to know whether hot Jupiters are indeed synchronously rotating \citep[e.g.,][]{Rauscher2014}.   

\begin{figure*}
\centering
\raisebox{5mm}{\includegraphics[width=39mm]{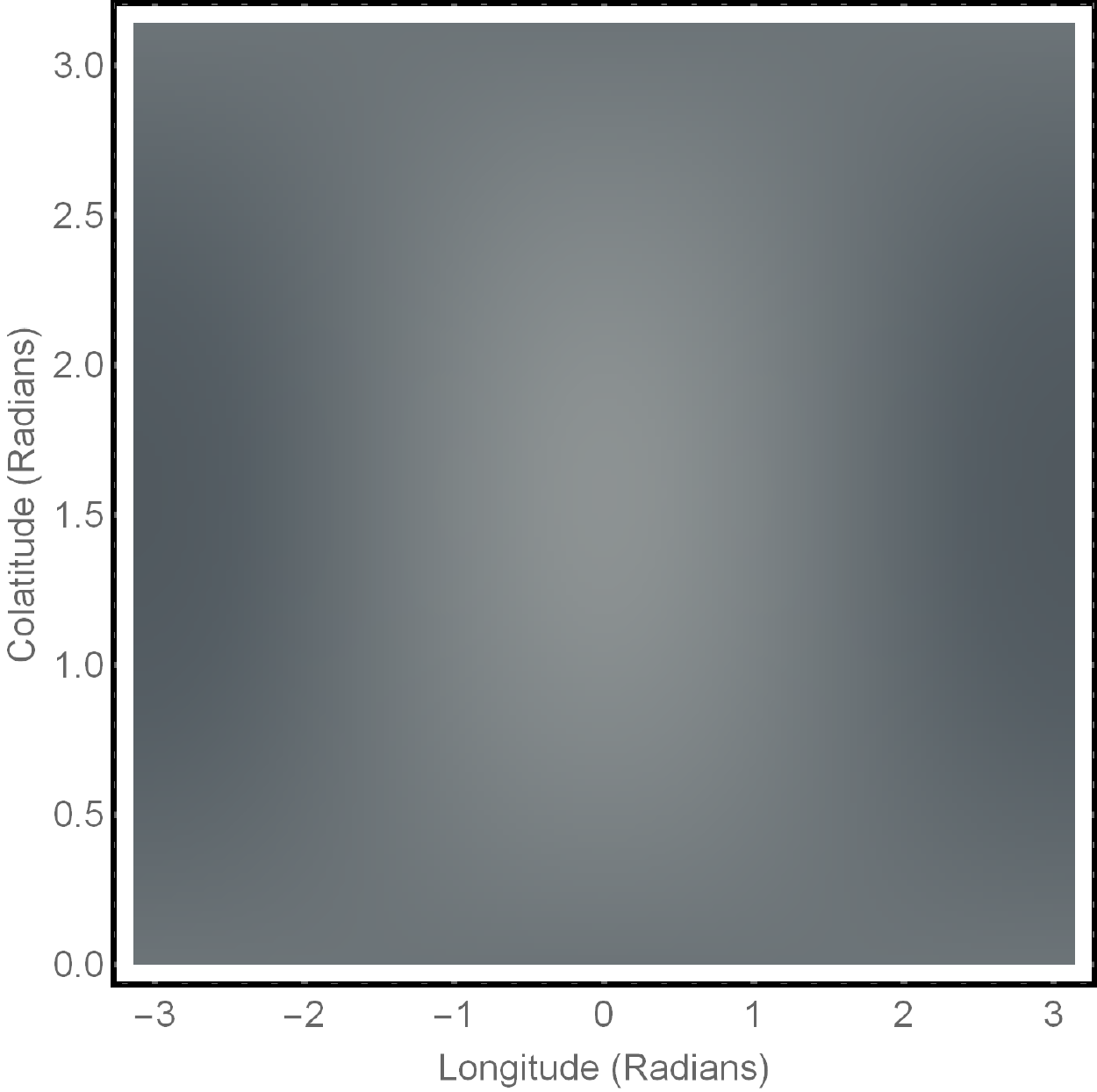}}
\raisebox{5mm}{\includegraphics[width=45mm]{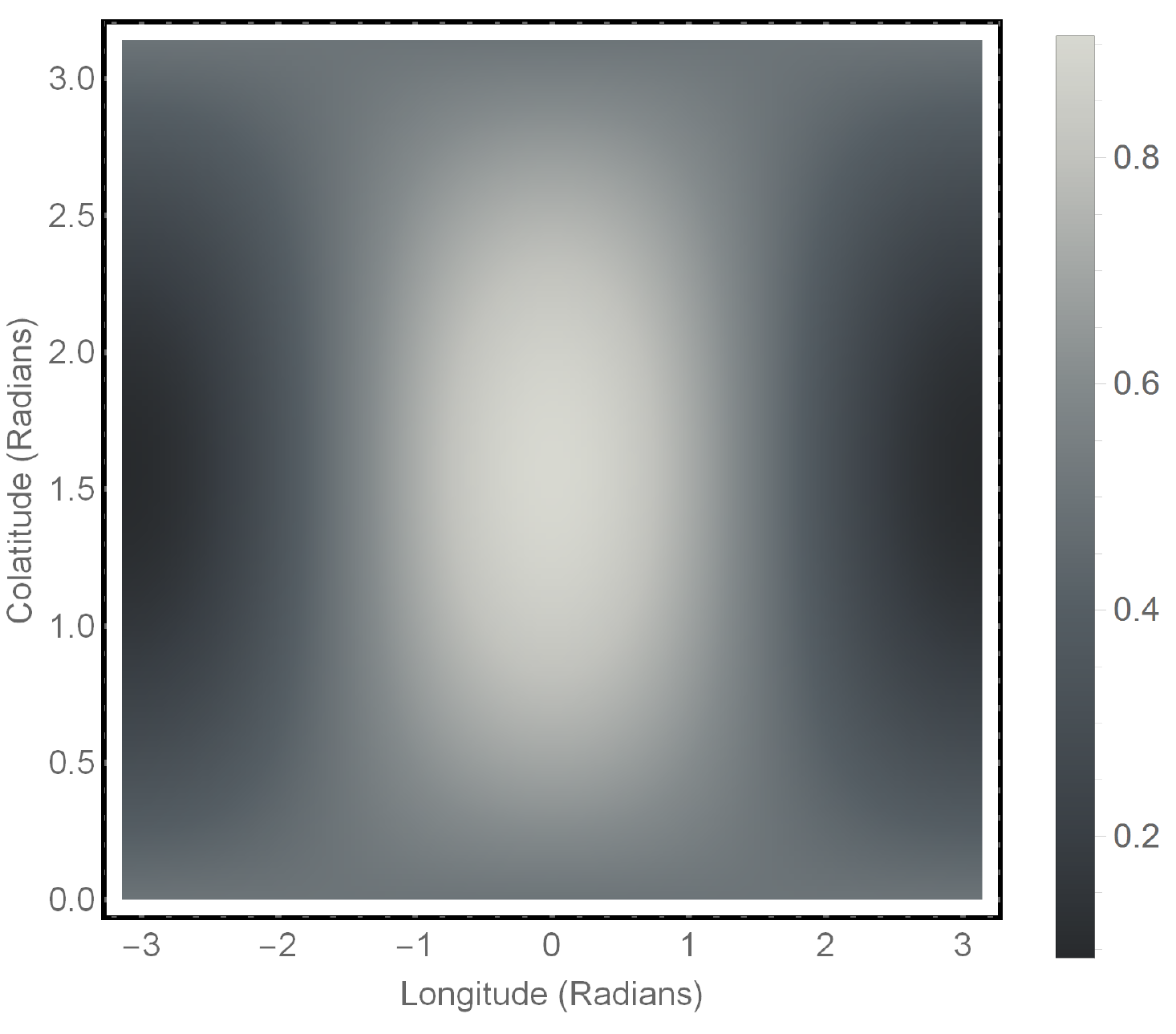}}
\raisebox{0mm}{\includegraphics[width=72mm]{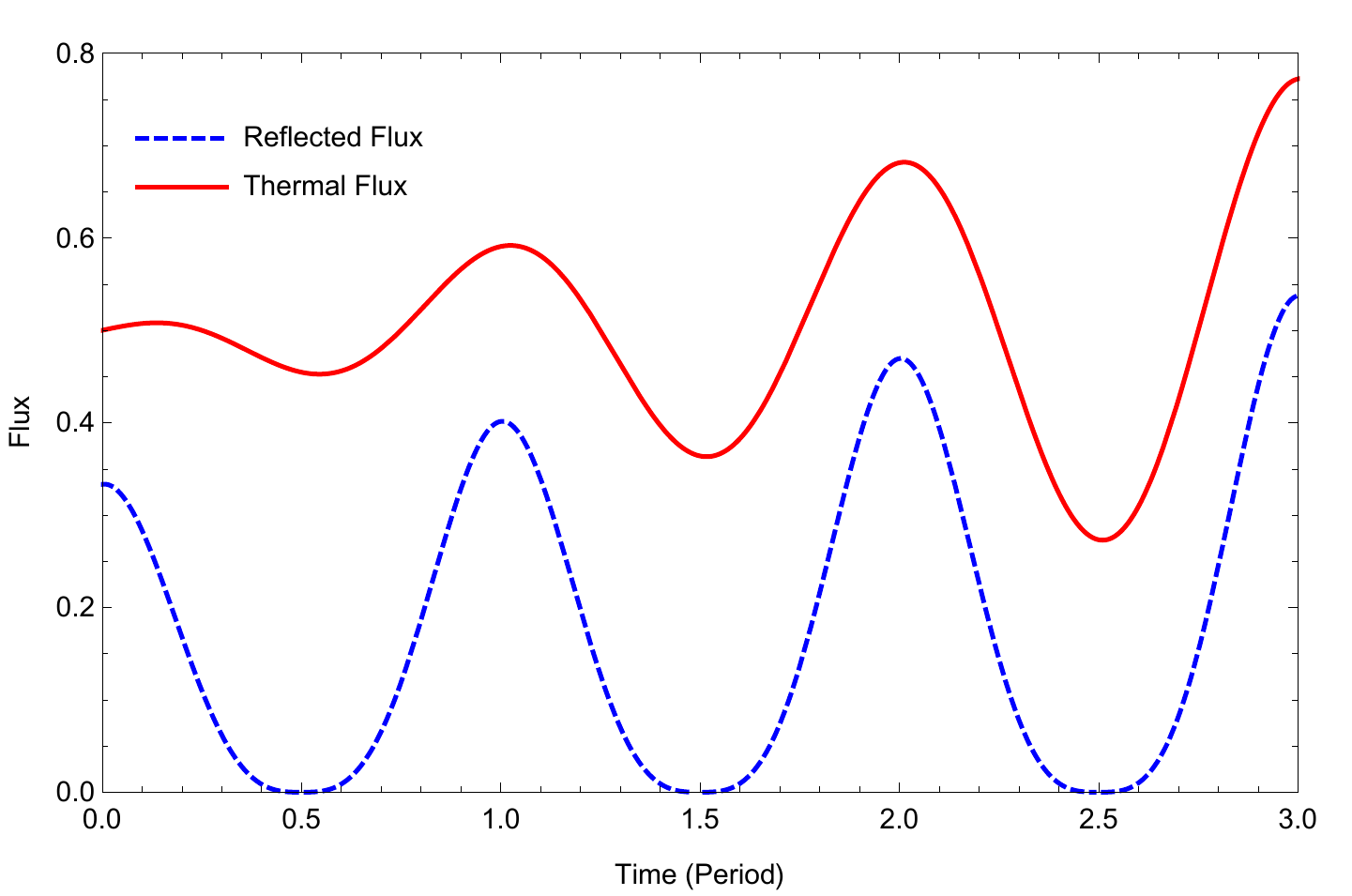}}
\caption{The planetary map at early (left) and late (right) times. The resulting thermal and reflected lightcurves for a zero-obliquity planet on an edge-on orbit are shown for three planetary orbits. The odd amplitude ratio for the thermal light curve is $A_3/A_0 = 0.022$, while for the reflective light curve it is $A_3/A_0 = 0.002$. For both thermal emission and reflected light, a time-variable map introduces power at odd modes in edge-on light curves. \label{time_variable}}
\end{figure*}
 
The map was taken to be $M(\Omega,t)=(1/2)Y_0^0-(t/80)Y_1^1(\Omega)$ for both thermal emission and reflected lightcurves (since we only report amplitude ratios, the unusual normalization is irrelevant). The coefficients were chosen so that the map is everywhere non-negative for the 3-orbit duration of observations (left panels of Figure~\ref{time_variable}). 

The thermal light curve corresponding to this map is $F(t)=(1/2)-\frac{t}{40 \sqrt{3}}\cos(\omega t)$ (right panel of Figure~\ref{time_variable}), where $\omega$ is the orbital frequency and the first three harmonic amplitudes are: $A_0=0.5$, $A_1=0.007$, $A_2=0.019$, and $A_3=0.011$. The relative size of the third harmonic is $A_3/A_0=0.022$, rather than zero for a static map.

The reflected light curve is given by $F(t)=\frac{1}{3\pi}\left(\sin\alpha +(\pi-\alpha)\cos(\alpha)\right)-\frac{\sqrt{3}}{160}t\cos^{4}(\omega t/2)$, where $\alpha$ is the orbital phase and the first three harmonic coefficients are: $A_0=0.135$, $A_1=0.167$, $A_2=0.030$, and $A_3=0.0003$. The relative size of the third harmonic is $A_3/A_0=0.002$. Again this is a substantial increase from the time independent case where the ratio is zero.

We chose a linear time dependence for simplicity, nonetheless it generated substantial third harmonics for both the thermal and reflected cases. The third harmonic could be maximized by choosing time dependence in the form of a square wave with frequency $3\omega$. While a more physically plausible model of time dependence should eventually be considered, our simple numerical experiment suggests that time variability in a planetary map can produce higher order harmonics---notably the $n=3$.

\section{The Presence of Odd Modes in Kepler Photometry}
A\&R15 reported high order harmonics for 16 Kepler Objects of Interest (KOI), 6 of which have since been deemed false positives.  
We downloaded all long-cadence (30~minute integration) data for the 10 remaining systems from MAST and closely follow the methodology of A\&R15: the flux was normalized to the mean for each quarter, the light curves were concatenated, and outliers were removed. 

Since A\&R15 used system parameters last updated on March 1, 2013, our results are slightly different from theirs. The last Vet update was September 18, 2015. Therefore, the transit duration, orbital period, time of the first transit, and every other parameter can be different (including the Kepler disposition, which is why 6 of these KOI are now categorized as false positives) .

\begin{figure}
\centering
\includegraphics[scale = 0.4]{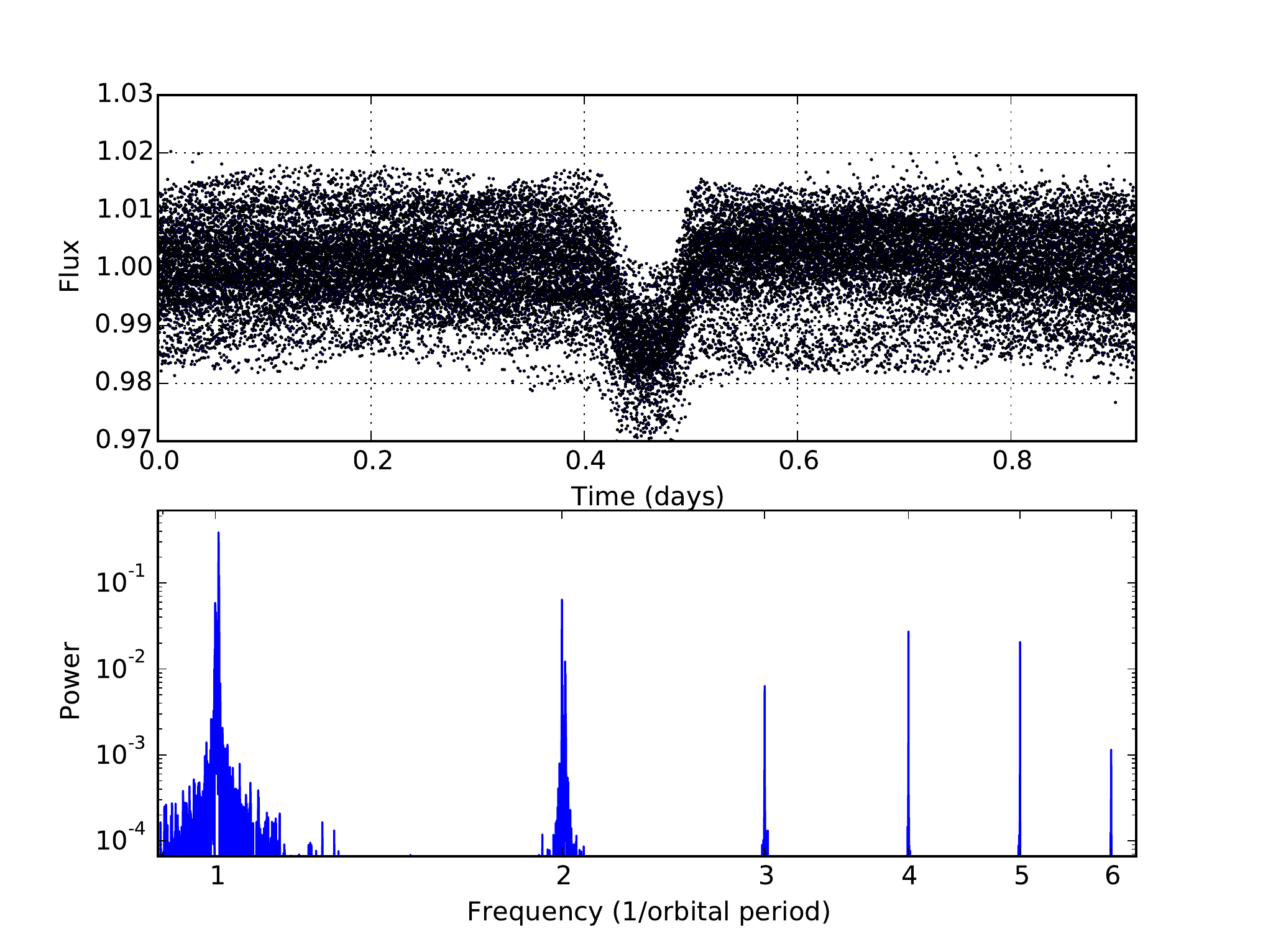}
\caption{Lomb-Scargle periodogram of K01546.01 using all the data, and computed with the AstroPy module.  Note the high power of odd harmonics; these are simply due to the periodic transit in the data. Only half the points of the lightcurve are plotted.}
\label{K01546.01_fulldata}
\end{figure}

Since transits and eclipses are periodic, a Lomb-Scargle (L-S) periodogram of the photometric time-series will exhibit peaks at the orbital frequency and its harmonics, as shown in Figure~\ref{K01546.01_fulldata}.  A\&R15 therefore removed the data within transits and eclipses, as has also been done by \cite{Shporer2015}. In fact, A\&R15 removed twice the predicted transit/eclipse duration, just in case there were transit timing or duration variations. 

\begin{figure}
\centering
\includegraphics[scale = 0.4]{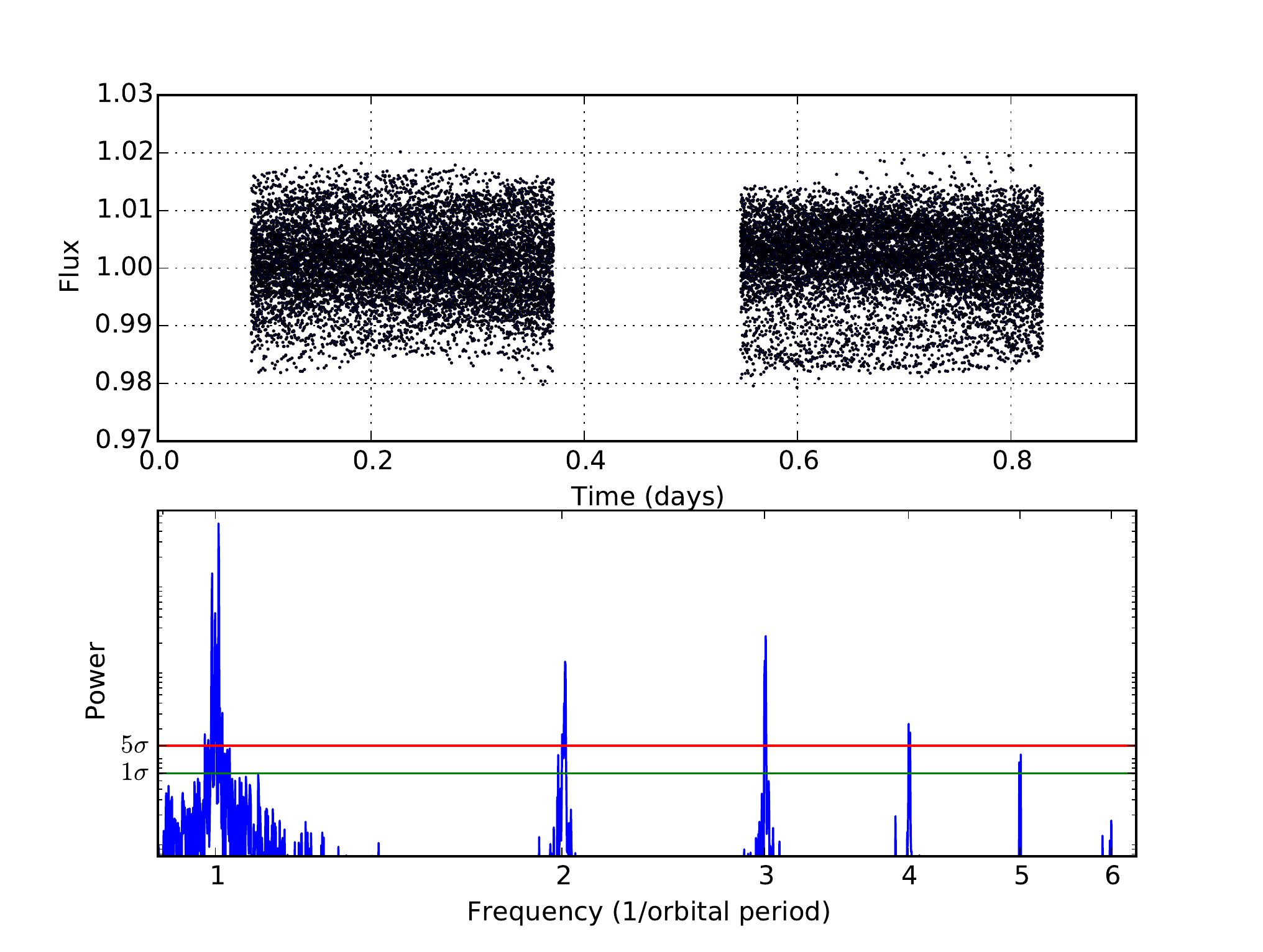}
\caption{\emph{Kepler} photometry of K01546.01 analyzed following the methodology of A\&R15 (cf.\ their Figures~1 and 2). The horizontal lines are the false alarm probabilities for 1$\sigma$ and 5$\sigma$. The L-S periodogram is plotted on a log-log scale. We can clearly see the high power in the 3rd harmonics. Only half the points of the lightcurves are shown.}
\label{K01546.01_K00823.01_AR}
\end{figure}

A\&R15 then computed the L-S periodogram of the remaining data using the SciPy module. The frequency was evenly sampled in log, ranging from $10^{-0.05}$ to $10^{0.08}$ times the orbital frequency, with 2000 frequency points. Figure~\ref{K01546.01_K00823.01_AR} shows that we can reproduce their results (c.f. Figure~1 of A\&R15). The small differences are most likely due to the fact that the MAST data have been updated since then. If one takes the false alarm probabilities at face value, then the high order harmonics, in particular $n=3$, are detected at very high significance.

\subsection{Periodogram Pitfalls}
There are several problems with the analysis of A\&R15. Firstly, 2000 frequencies are insufficient to resolve the harmonic peaks. For all of our plots except for Figure~\ref{K01546.01_K00823.01_AR} we therefore use the new Lomb-Scargle module from AstroPy, which automatically determines how many frequencies are needed. (This module also correctly manages uncorrelated Gaussian measurement error, and is faster: O[$N\log N$] instead of O[$N^2$].)

Secondly, false alarm probabilities for L-S Periodograms are not robust because they assume white noise. This is a problem because the \emph{Kepler} photometry for many of these targets exhibits clear astrophysical variations that are unmodeled and therefore represent time-structured noise; astrophysical processes that are properly modeled do not impact the L-S periodogram of the residuals, as described below. 

Lastly, the gaps created by removing the transits and eclipses significantly impact the L-S periodogram.  

\subsection{Numerical Experiments}
\begin{figure}
\centering
\includegraphics[scale = 0.4]{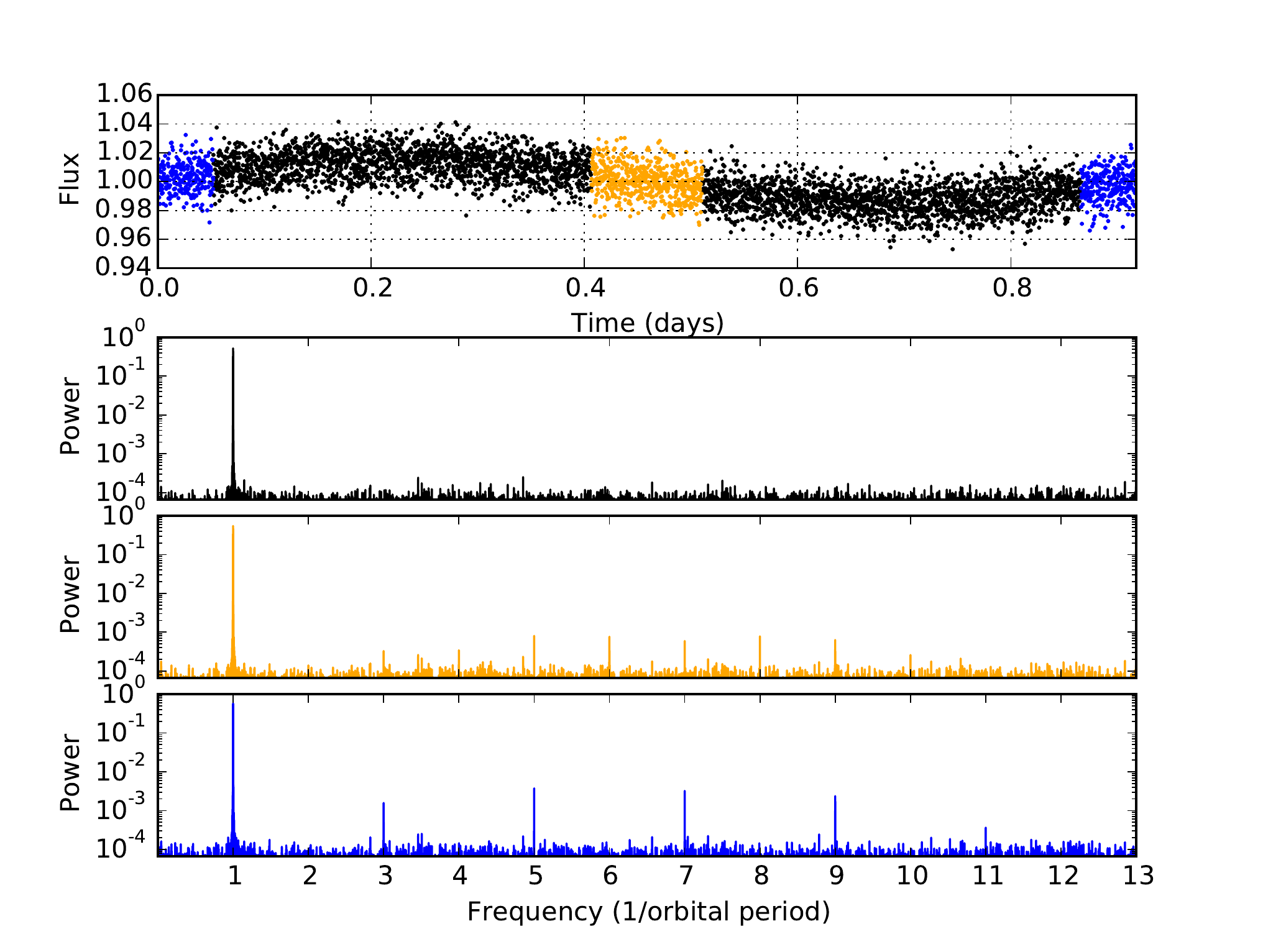}
\caption{These data simulate the residuals after modeling out transits and eclipses of a planet---but not stellar variability. They were generated using the following parameters: 1000 days of 30-minute cadence photometry (only 1/10th of the points are plotted), 0.917569588 day orbital period, and the phase amplitude (reflected light, thermal emission, or Doppler beaming) is \num{0.015}. The noise standard deviation is 0.01. The removal of transits and eclipses is simulated by removing a 2.5 hour chunk of data every 0.917569588 days.  The black L-S periodogram  was done on the whole data set, the orange used the data after the transit has been removed and finally, the blue one was done on the data after the transit and the eclipse have been removed. The L-S power is on a log scale and the frequency is on a linear scale. We can clearly see that removing the transit adds power to the high order harmonics and that removing both the transit and the eclipse boosts by orders of magnitude the power of the odd harmonics.}
\label{artificial_data}
\end{figure}

We generate synthetic data to demonstrate the problem with removing transits rather than modeling them. Figure~\ref{artificial_data} shows what happens with a simple sinusoid: removing transits introduces peaks at the orbital harmonics, while additionally removing eclipses enhances the odd harmonics. 

We also tested what happens if periodic chunks of data are removed from a time series with Gaussian noise but no signal: the L-S periodogram did not exhibit power in orbital harmonics. We surmise that removing periodic chunks of data only affects the periodogram if the data already contain periodic signals.

We performed a few more tests on synthetic data. Firstly, a signal of twice the original sinusoid frequency was added to simulate ellipsoidal variations in addition to the fundamental mode. If all the data are kept, then the only peaks in the periodogram were at 1 and 2 times the orbital frequency, as one would expect. If only the transit is removed, then small peaks appear at the harmonics. However, if the eclipse is removed as well, then significant peaks appear at every harmonic. The height of those peaks depends upon the relative amplitude of the two sinusoids.

Secondly, we tried to simulate stellar variability in the absence of planetary phase variations. The input sinusoid now represents stellar variations due to star spots rotating in and out of view and the frequency at which data are cut is the exoplanet orbital frequency (an irrational multiple of the stellar rotation). Basically, there's a sine wave at a certain frequency but chunks of data are removed at a different frequency. If no data are removed, then the only peak is of course at the stellar frequency. If the transits and the eclipses are removed, then there are additional peaks at $f = (f_{\rm star} + 2n)*f_{\rm orb}$, where $f$ is the frequency of the peaks, $f_{\rm star}$ is the stellar frequency, $f_{\rm orb}$ is the orbital frequency and $n = \{-1, 0, 1, 2, 3\}$. Notably, this scenario does not produce peaks at the orbital harmonics of the planet. 

Lastly, if we do the same experiment as above but this time we add a sinusoid at the orbital frequency to the stellar variations' sinusoid, then we observe new peaks at the orbital frequency's odd harmonics in addition to the ones that were there without this extra sine. The power of these peaks depends on the relative amplitudes of the sinusoids.

To summarize the results of our numerical experiments: periodic gaps at the fundamental frequency or its harmonics conspire with true signal at the fundamental mode to produce spurious periodogram peaks at the orbital harmonics. This is an important cautionary tale, as L-S periodograms are used instead of Fourier transforms precisely because of gaps in data.

\subsection{Model Occultations, Don't Remove Them}
\begin{figure}
\centering
\includegraphics[scale = 0.4]{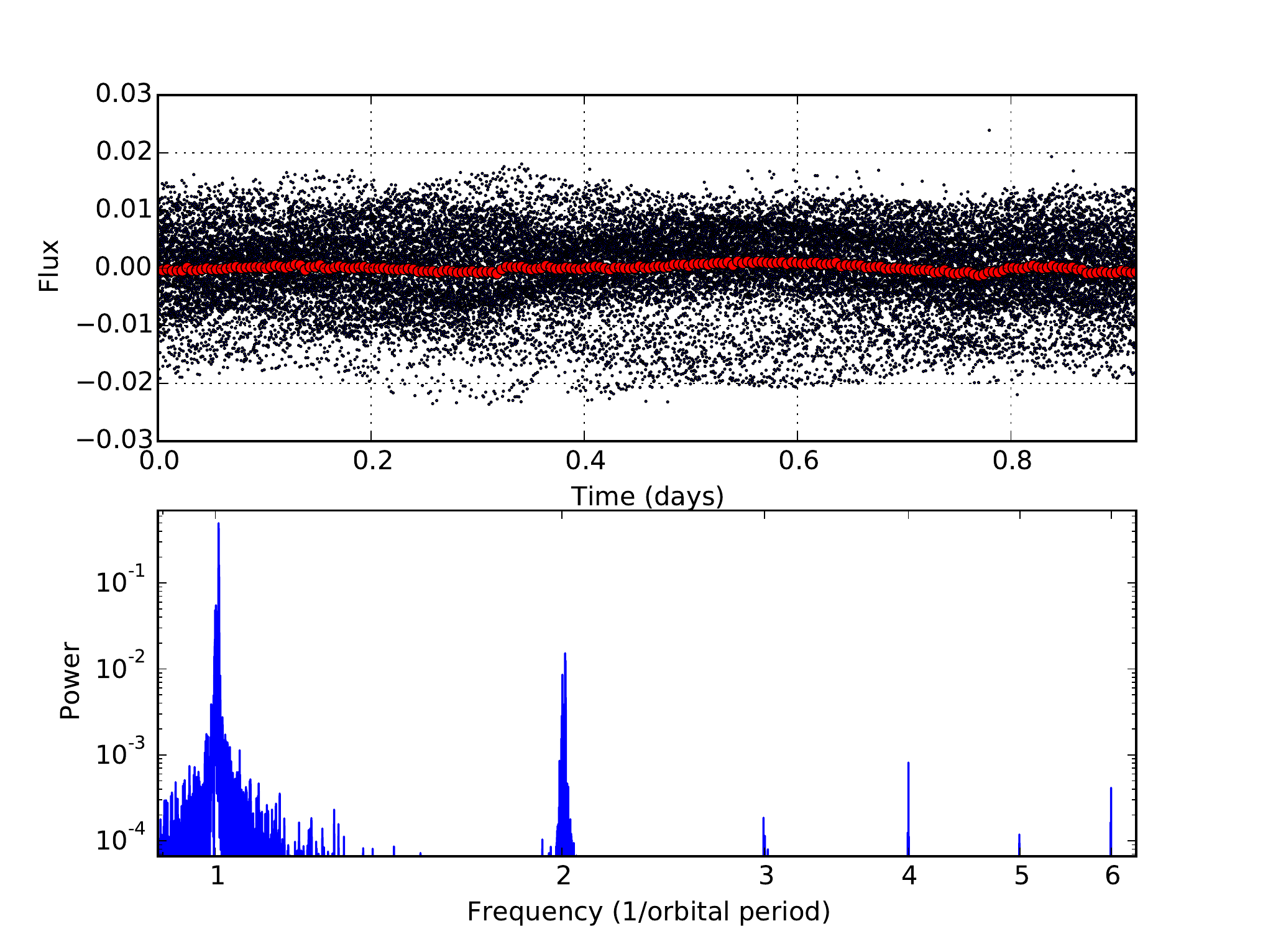}
\caption{Periodograms for K01546.01. The upper panel now shows residuals. The L-S periodogram is shown on a log-log plot. Peaks at m=3,4,5 and 6 are visible, but orders of magnitude smaller than in Figure~\ref{K01546.01_K00823.01_AR}. Only half the points of the lightcurves are plotted. The red points are the binned data.}
\label{K00256.01andK00823.01_Elie}
\end{figure}

Keeping the transits in the lightcurve or removing those data both lead to high-order harmonics that have nothing to do with the planet's phase variations. We therefore \emph{model} the transits following \cite{Rowe2015}, divide out the lightcurves by that model, then compute the L-S periodogram of the residuals.\footnote{Ideally, one would run a simultaneous fit for all of the relevant astrophysics.  In practice, this approach is complicated because inhomogeoneous clouds, weather, star spots, and planet-star interactions are difficult to parameterize \citep[e.g., fitting star-spot models to long-baseline Kepler photometry of \emph{isolated stars} can prove dicey:][]{Aigrain2015}.  We therefore limit ourselves to fitting the---easily parameterized---transits and computing the periodogram of the residuals.} As shown in Fig. \ref{K00256.01andK00823.01_Elie}, the power in the high order harmonics are decreased by orders of magnitude.

Since the false-alarm probabilities returned by L-S codes are not trustworthy, we estimate the uncertainty on the spectrum in two residual permutation methods: bootstrap and cross-validation.

\begin{figure}
\centering
\includegraphics[scale = 0.4]{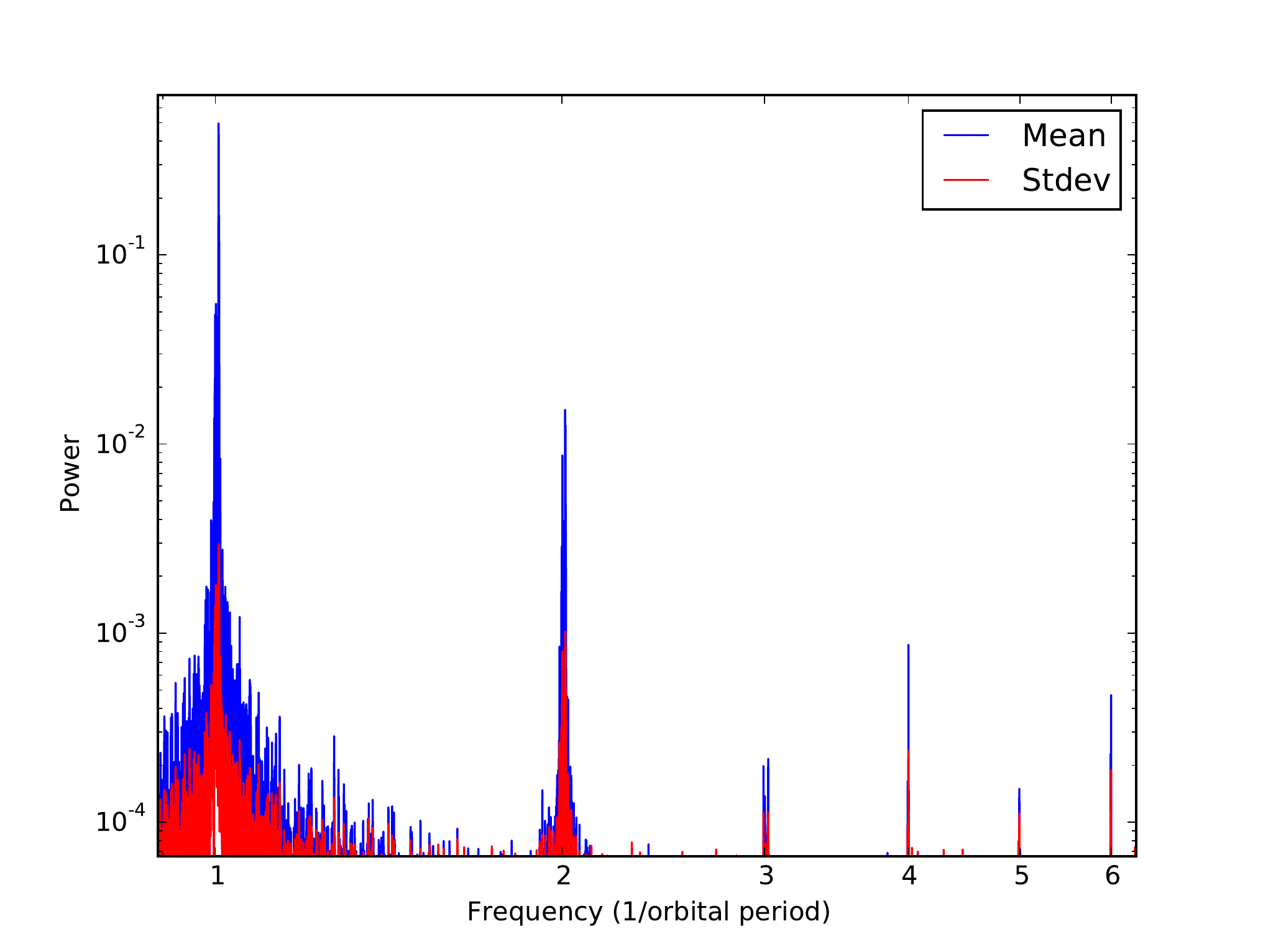}
\caption{Bootstrap uncertainty analysis for the K01546.01 L-S periodogram (see text for details). One cannot assign a false-alarm probability that is independent of frequency as is often assumed in L-S analyses.  The red line can loosely be thought of as the 68\% false-alarm \emph{spectrum}. Although the first and second peaks (fundamental mode and first harmonic) are robust, the higher-order modes are only detected at $\lesssim 2\sigma$. Our cross-validation, also described in the text, yields a higher noise spectrum and hence lower significance for the odd harmonics.}
\label{bootstrap_0.5}
\end{figure}

For the bootstrap uncertainty analysis, 50$\%$ of the data were randomly removed. A L-S was then done on the remaining data points. This process was done 50 times, each time on the same frequency sample. Then, using the 50 powers per frequency obtained, we computed the mean and the standard deviation at each frequency, shown in Figure~\ref{bootstrap_0.5}. The only significant modes are the fundamental and the first harmonics. 

For the cross-validation, 50$\%$ of the data were again randomly chosen and their L-S periodogram was computed. Another L-S periodogram was also computed on the omitted data. The absolute difference between the computed powers was calculated for each frequency. This process was repeated 50 times. Afterwards, we computed the mean of the absolute difference in power at each frequency, as an estimate of the uncertainty spectrum. The uncertainty spectrum from cross-validation (not shown) is even greater than that from the bootstrap (Figure~\ref{bootstrap_0.5}), reinforcing our conclusion that the higher-order harmonics are not robust.

Although we have only presented plots for K01546.01, our results are qualitatively the same for all of the vetted A\&R15 systems, with a few exceptions described below. We did not consider K02867.02, because its transit depth is only 20.3 ppm according to MAST, so it is unlikely that this exoplanet would exhibit detectable phase curve variations. 

At the other end of the signal-to-noise spectrum, HAT-P-7b and Kepler-13Ab (a.k.a.\ K00013.01) both have significantly asymmetric transits, likely due to planet-induced gravity darkening of the star \citep[][]{Morris2013}. Using another model to sidestep the transit asymmetry, \cite{Esteves2015} showed that these exoplanets exhibit power in the third harmonic. We did not attempt to replicate their approach, but we discuss these planets in Section~6.3.\footnote{If one fits these planets with a symmetric transit then there are significant residuals in transit.  Roughly speaking, this can be thought of as a delta function at the orbital frequency.  The resulting Fourier spectrum exhibits peaks at all harmonics, masking any harmonics due to phase variations.}

\section{Discussion}
\subsection{Weather on Other Worlds}
Although there is ample evidence of time-evolving clouds on brown dwarfs \citep{Artigau2009, Radigan2012, Radigan2014, Metchev2015}, there is minimal evidence for time-variability in the atmospheres of short-period planets. The most constraining searches for time-variable emission from exoplanets have consisted of repeated eclipses measured with Spitzer/IRAC \citep{Werner2004,Fazio2004}. \cite{Agol2010} studied the transiting planet system HD 189733 at 8 micron, establishing a 1$\sigma$ upper limit on variability of 2.7\% of day-side planetary flux changes and 1$\sigma$ upper limit on variability of 17\% on the night-side planetary flux. \cite{Knutson2011} established a 1$\sigma$ limit of 17\% on changes in extrasolar planet GJ 436b's dayside flux in the 8 micron band.  \cite{Wong2014} determined a 1$\sigma$ upper limit on dayside 4.5~micron variability of Hot Jupiter XO-3b of 5\%. \cite{Demory2016} report a 4$\sigma$ detection of variability in the dayside thermal emission from the transiting super-Earth 55 Cancri e, but the reported variability is at the accuracy limit of the Spitzer Space Telescope \citep{Ingalls2016}.

We therefore expect brown dwarf rotational light curves to contain appreciable $A_3$, both because their maps change with time.\footnote{Brown Dwarfs also have isotropic inclinations and may have N-S asymmetric maps, which would produce odd harmonics in their light curves (cf.\ Section~3.1).}  We do not, however, believe that hot Jupiter variability is sufficiently large to contribute significant $A_3$ to their orbital light curves: the extreme scenario we adopted in Section~4---order unity change in the map over the course of a few orbits---produces odd power that is a factor of 3--9 too small to account for the measured signals in Kepler-13Ab and HAT-P-7b.  

\subsection{Other Astrophysical Sources of Odd Harmonics}
Doppler beaming and ellipsoidal variations are often modeled as purely sinusoidal at the planet's orbital frequency and its first harmonic, respectively.  This is exact for the Doppler beaming since the radial velocity of the star is indeed exactly sinusoidal for a circular orbit.  Ellipsoidal variations are more subtle, however.

Ellipsoidal variations are primarily due to the phase-dependent projected area of the star and planet \citep[e.g.,][]{Cowan2012}:
\begin{equation}\label{ellipsoidal}
A_{p}(\alpha)=\pi R^2_p\sqrt{\cos^2\alpha+\frac{R_{\text{long}}}{R_p}\sin^2\alpha},
\end{equation}
where $R_p$ is the planetary radius perpendicular to the star--planet axis, $R_{\rm long}$ is the planet's radius along the star--planet axis, and $\alpha$ is the usual orbital phase: $\alpha=0$ at eclipse, $\alpha=\pi/2$ at quadrature. 

Equation~\ref{ellipsoidal} satisfies $f(x+\pi)=f(x)$, and thus by Theorem 1 of Appendix \ref{TheAppendix}, can produce no odd harmonics.

In practice, however, higher-order effects kick in, such that the body is not perfectly ellipsoidal, nor does it have uniform intensity.  These subtleties contribute to $A_1$ and $A_3$, in addition to $A_2$ \citep{Morris1985}.  In particular, the $A_3$ due to tides is proportional to $(R_*/a)^4$, where $R_*$ is the stellar radius and $a$ is the orbital semi-major axis. For HAT-P-7b and Kepler-13Ab, the figures of merit for tidal odd modes are $3\times10^{-3}$ and $2\times10^{-3}$, respectively.

\subsection{The Origin of Odd Modes}
Even after modeling out the transits, we find non-zero odd-harmonics for all of the vetted KOI studied by A\&R15.  But the amplitudes we measure are at least an order of magnitude lower than that reported by A\&R15 because we model transits and eclipses rather than removing them. Moreover, our uncertainty estimates suggest that the $A_3$ are insignificant (e.g., less than 3$\sigma$ for KOI1546.01). Nonetheless, \cite{Esteves2013, Esteves2015} and \cite{Shporer2014} reported 10--20$\sigma$ detections of  $A_3$ of HAT-P-7b and Kepler-13Ab, even after modeling transits and eclipses. Esteves et al.\ hypothesized that the spin-orbit misalignment of the stars enhanced the third harmonic contribution of ellipsoidal variations \citep[it is possible that both planets are on retrograde, nearly polar orbits around their respective stars][]{Narita2009, Winn2009, Barnes2011, Szabo2011}. \cite{Esteves2015} further noted that the third mode persists during eclipse---when the planet is hidden behind its star---suggesting that it is due to variations in starlight rather than planetary light.

While we express Fourier amplitudes in terms of the orbit-averaged planetary flux, $A_3/A_0$, observers often express amplitudes in terms of the \emph{stellar} flux, $A_3/F_*$.   In order to translate between the two conventions, one needs to know the planetary flux relative to its host star.  For eclipsing planets, $A_0/F_*$ can be directly measured because we get to observe the star when the planet is hidden from view (i.e., during eclipse). For non-eclipsing systems $A_0$ cannot be directly measured,\footnote{Long-period non-eclipsing planets can be directly imaged to measure their orbit-averaged flux, but such planets are outside the scope of this study.} but can be roughly estimated via the thermal or reflected figures of merit (FoM):
\begin{eqnarray}
{\rm FoM}_{\rm therm}(\lambda) = &\frac{B(T_{\rm eq}, \lambda)}{B(T_*, \lambda)} \left(\frac{R_p}{R_*}\right)^2\\
{\rm FoM}_{\rm refl} = &\left(\frac{R_p}{a}\right)^2,
\end{eqnarray}
 where $B(T, \lambda)$ is the Planck function for a blackbody of temperature $T$, and wavelength $\lambda$, $T_{\rm eq}$ and $T_*$ are the planet's equilibrium temperature and the star's brightness temperature, $R_p$ is here the mean planetary radius. Note that the thermal figure of merit is a function of wavelength, but the reflected light figure of merit is not.
 
The thermal figures of merit for HAT-P-7b and Kepler-13Ab in the Kepler band are $1\times10^{-5}$ and $7\times10^{-5}$, respectively. Multiplying these numbers by the \emph{maximum} possible thermal ratio $A_3/A_0 \approx 10^{-2}$, we find that N-S asymmetry in planetary emission is more than an order of magnitude too weak to explain the measured third mode amplitude of 2--7$\times$10$^{-6}$.  

The reflected light figures of merit for HAT-P-7b and Kepler-13Ab are $3\times10^{-4}$ and $4\times10^{-4}$, respectively. Depending on the optical geometric albedo of the planet \citep[0.01--0.30;][]{Dai2016, Demory2011}, we therefore conclude that the contribution from inhomogeneous planetary albedo is also at best an order of magnitude too small to account for the reported signals of 2--7$\times$10$^{-6}$.  

While time-variable thermal emission or albedo could, in principle, produce variations of this scale, it would require planet-scale variations at the third harmonic of the orbital frequency, which seems contrived and implausible.

We therefore favor---like Esteves et al.---the tidal hypothesis: the planet raises tides on its host star, which in turn cause the star to have a non-uniform surface temperature.  We urge stellar theorists to model this scenario, and observers to acquire orbital light curves of these systems at different wavelengths (e.g., with the James Webb Space Telescope).   

\section{Conclusions}
Many authors have attributed each Fourier coefficient in a phase curve to a distinct astrophysical cause \citep{Welsh2010, Faigler2011, Faigler2013, Faigler2015, Barclay2012, Shporer2011, Shporer2015, Jackson2012, Angerhausen2015, Armstrong2015}. Although this approach is suitable to search for non-transiting planets or obtain rough estimates of planetary mass, using over-simplified astrophysical models may lead to imperfect fits and therefore spurious claims of new physics. 

Only \cite{Esteves2013, Esteves2015} used both a Lambert phase curve and the full formulation of ellipsoidal variations \citep{Morris1985}.  Even the Esteves et al.\ formulation does not fully account for the effects of inhomogeneous albedo; they merely allow for a phase offset in the Lambert phase curve, which is neither complete, nor self-consistent.  

Our foray into the analysis of \emph{Kepler} light curves has taught us that if you suspect an astrophysical signal to be present in the data---stellar variability, transits, reflected light, or ellipsoidal variations---then it behooves you to model it properly before moving on to higher order effects. Since many astrophysical processes can produce similar signals, one must simultaneously fit for all of the parameters in order to make robust inferences about specific astrophysical causes. 

Moreover, removing periodic chunks of data before performing a Lomb-Scargle periodogram is only acceptable if your residuals are Gaussian---but if your residuals are Gaussian you are unlikely to run a L-S.  So just don't cut out data and beware of running L-S on data with regular gaps. 

To first order, transiting planets cannot contribute odd harmonics to the phase variations of the system.  However, we have seen that North-South asymmetric maps and/or  time variable maps can produce small amplitude odd harmonics.  Curiously, in either case the odd modes would be due to weather on an exoplanet: short-period planets are expected to have zero obliquity and hence N-S symmetric stellar forcing.  North-South asymmetry in a giant planet would therefore suggest stochastic localized features, such as weather. The theoretical maximum for these contributions is at the detection limit for existing Kepler photometry for the brightest targets, but may be detected or usefully constrained with future space telescopes. 

The odd harmonic contribution of tidal effects should drop off more steeply with semi-major axis than the reflected light of exoplanets: $a^{-4}$ vs.\ $a^{-2}$. Moreover, short-period planets are expected to be tidally locked even if they do not noticeably tidally warp their host star \citep{Peale1977}.  We therefore expect that slightly cooler planets---orbiting further from somewhat smaller stars---may betray their weather via odd harmonics.

\section*{Acknowledgments}

We thank Caden Armstrong and Hanno Rein for sharing unpublished data products, as well as for discussing details of their---and our---analysis.  We also thank Avi Shporer for useful discussions. We thank Daniel Angerhausen and Emerson DeLarme for sharing ellipsoidal coefficients. \'Elie Bouffard was supported by an iREx summer internship. We thank the International Space Science Institute in Bern, Switzerland, for hosting the Exo-Cartography workshop series.

%%%%%%%%%%%%%%%%%%%%%%%%%%%%%%%%%%%%%%%%%%%%%%%%%%



%%%%%%%%%%%%%%%%%%%%%%%%%%%%%%%%%%%%%%%%%%%%%%%%%%

%%%%%%%%%%%%%%%%% APPENDICES %%%%%%%%%%%%%%%%%%%%%

\onecolumn
\appendix

\section{Mathematical Proof of No Odd Harmonics in the Reflected Edge-On Case}
\label{TheAppendix}

In this section we present an analytic proof that there can be no odd harmonics in the $n>1$ case for edge on reflected light curves. For the reflective light curve, the kernel dependends on both visibility and illumination. However, it simplifies in the tidally locked, edge-on case to: $K(\theta , \phi , t) = \frac{1}{\pi}\sin^2 \theta\cos\phi \cos ( \phi - \phi_{o})$, where $\phi_o$ is the sub-observer longitude. Then the flux is given by 
\begin{equation}
	F_l^m(t)  =  \frac{N_l^m}{\pi} \int_{-1}^{1}(1-x^2)P_{lm}(x) dx  \int_{\phi_1}^{\phi_2} \cos\phi \cos(\phi-\phi_o) \cos(m\phi) d\phi \equiv \frac{N_l^m}{\pi} \int_{-1}^{1}(1-x^2)P_{lm}(x) dx \ \Phi_{m}(\phi_{o}) ,
\end{equation} 
where $\phi_1 = \max[-\pi/2, \phi_{o}-\pi/2]$, $\phi_2 = \min[\pi/2, \phi_{o}+\pi/2]$, and the last equality defines the shorthand $\Phi_{m}(\phi_o) = \int_{\phi_1}^{\phi_2} \cos\phi \cos(\phi-\phi_o) \cos(m\phi) d\phi$. From \cite{Cowan2013}, the solutions for the $\Phi_{m}(\phi_o)$ functions are:

\begin{equation}\label{edge_on_reflected_phi_integral}
		\Phi_m(\phi_o)  = \left\{ \begin{array}{ll} \frac{-\sin(m\phi_o/2)}{m(m^2-4)}\Big( (m+2) \sin\left(\alpha - \frac{m\alpha}{2} +\frac{m\pi}{2}\right) + (m-2) \sin\left(\alpha + \frac{m\alpha}{2} -\frac{m\pi}{2}\right)\Big) & \textrm{if $m<-2$}\\[6pt]
			\frac{1}{4}\sin\phi_o(\pi-\alpha+\sin\alpha\cos\alpha) & \textrm{if $m=-2$}\\[6pt]
			\frac{1}{3}\sin\phi_o(1+\cos\alpha) & \textrm{if $m=-1$}\\[6pt]
			\frac{1}{2}\Big(\sin\alpha+(\pi-\alpha)\cos\alpha\Big) & \textrm{if $m=0$}\\[6pt]
			\frac{4}{3}\cos^4(\phi_o/2) & \textrm{if $m=1$}\\[6pt]
			\frac{1}{4}\cos\phi_o(\pi-\alpha+\sin\alpha\cos\alpha) & \textrm{if $m=2$}\\[6pt]
			\frac{\cos(m\phi_o/2)}{m(m^2-4)}\Big( (m+2) \sin\left(\alpha - \frac{m\alpha}{2} +\frac{m\pi}{2}\right) + (m-2) \sin\left(\alpha + \frac{m\alpha}{2} -\frac{m\pi}{2}\right)\Big) & \textrm{if $m>2$}. \end{array} \right.
\end{equation}

Note that $\alpha$ is $|\phi_o|$. Furthermore, $\alpha$ is only defined in the domain $[-\pi, \pi]$; because of this restriction, we must be careful in the techniques we use to analyze the light curves.

We begin by recalling a few facts from the foundations of Fourier theory. If $f(x)$ is an even function, $f(-x)=f(x)$, then its Fourier expansion will consist only of cosine terms. The $n$th Fourier coefficient is therefore found by integrating \[
			a_n=\frac{1}{\pi}\int_{-\pi}^{\pi}f(x)\cos(nx)dx.
			\]
If $f(x)$ is an odd function, $f(-x) = -f(x)$, then its Fourier expansion will consist only of sine terms. The $n$th Fourier coefficient is then
    \[
	b_n=\frac{1}{\pi}\int_{-\pi}^{\pi}f(x)\sin(nx)dx.
	\]
Finally, the Fourier Series of the sum of two functions $f+g$ is equal to the Fourier series of $f$ added to the Fourier series of $g$. This holds for any finite sum of functions.\\

\noindent
\textbf{Theorem 1}: If a function $f$ satisfies $f(x+\pi)=f(x)$, then there are no odd harmonics beyond the fundamental mode.
	
\noindent
\textbf{Proof:} We first consider the cosine terms of a function that satisfies $f(x+\pi)=f(x)$. Note that the odd harmonic coefficients $a_{2n+1}$ are found via integrating \[
a_{2n+1}=\frac{1}{\pi}\int_{-\pi}^{\pi}f(x)\cos((2n+1)x) dx.
\]We substitute $x+\pi$ for $x$:
\begin{align*}
	&\frac{1}{\pi}\int_{0}^{2\pi}f(x+\pi)\cos((2n+1)(x+\pi)) dx  = \frac{1}{\pi}\int_{0}^{2\pi}f(x)\cos((2n+1)(x+\pi)) dx \\
	&\;\;\;=\frac{1}{\pi}\int_{0}^{2\pi}f(x)\cos((2n+1)x+\pi) dx =-\frac{1}{\pi}\int_{0}^{2\pi}f(x)\cos((2n+1)x) dx.
\end{align*}
This implies $a_{2n+1}=-a_{2n+1}$, so it follows that $a_{2n+1}=0$ for all $n\in\mathbb{N}$. A structurally identical argument can be used for the odd harmonic sine coefficients.

\noindent
\textbf{Lemma 1.1}: If an even function $f$ defined only on the domain $[-\pi, \pi]$ satisfies $f(x+\pi)=f(x)$, or $f(\pi-x)=f(x)$, then there can be no odd harmonics beyond the fundamental mode. 

\noindent
\textbf{Proof:} Suppose $f$ satisfies $f(x+\pi)=f(x)$. The odd harmonic coefficients are found via integrating\[
a_{2n+1}=\frac{1}{\pi}\int_{-\pi}^{\pi}f(x)\cos((2n+1)x) dx.
\]However, as $f$ is even, we can equivalently integrate \[
a_{2n+1}=\frac{2}{\pi}\int_{-\pi}^{0}f(x)\cos((2n+1)x) dx.
\]
We substitute $x+\pi$ for $x$:
\begin{align*}
	&\frac{2}{\pi}\int_{0}^{\pi}f(x+\pi)\cos((2n+1)(x+\pi)) dx  = \frac{2}{\pi}\int_{0}^{\pi}f(x)\cos((2n+1)(x+\pi)) dx \\
	&\;\;\;=\frac{2}{\pi}\int_{0}^{\pi}f(x)\cos((2n+1)x+\pi) dx =-\frac{2}{\pi}\int_{0}^{\pi}f(x)\cos((2n+1)x) dx.
\end{align*}
As this implies $a_{2n+1}=-a_{2n+1}$, it follows that $a_{2n+1}=0$ for all $n\in\mathbb{N}$. A structurally similar argument can be made for even functions of the form $f(\pi-x)=f(x)$ on the domain $[-\pi, \pi]$.

Throughout the remainder of this appendix the strategy will be to transform the trigonometric expressions from Eq. \eqref{edge_on_reflected_phi_integral} until we can either apply Lemma 1.1 to claim that they do not contribute odd harmonics, or until we can show this explicitly. In all cases both techniques will be needed for various contributing terms. 

To illustrate the technique we begin with the $m=0$ case. In this case Eq. \eqref{edge_on_reflected_phi_integral} tells us that 
\[
\Phi_{0}(\phi_o) = \frac{1}{2}[\sin \alpha +(\pi - \alpha)\cos \alpha],
\]
where, recall, that $\alpha = |\phi_o|$. Due to the absolute value contained in the argument the first term, $\frac{1}{2} \sin(\alpha)$ is an even function, and satisfies $f(\pi-x)=f(x)$, thus by Lemma 1.1 it cannot contribute any odd order terms for $n>1$. Next, because cosine is even, note that $\frac{\pi}{2}\cos(\alpha)=\frac{\pi}{2}\cos(\phi_o)$ and hence can only contribute to the fundamental mode.  Finally, we integrate the last term $-\frac{1}{2}\alpha\cos(\alpha)$ directly for its cosine coefficients:
\[
-\frac{1}{\pi}\int_{0}^{\pi}\phi_o\cos(\phi_o)\cos(n\phi_o)d\phi_o=\frac{1+n^2+(1+n^2)\cos(n\pi)}{(-1+n^2)^2\pi}.
\]
For odd $n>1$, the right hand side vanishes and hence $\alpha\cos(\phi_o)$ cannot contribute any odd order terms beyond the fundamental. For the fundamental, $n=1$, the denominator is 0 and the formula breaks down; an independent calculation shows that this coefficient does not vanish, but as this is the fundamental it is not of interest. This completes the demonstration that there are no odd harmonic contributions to the harmonic light curve for $m=0$.

In the cases with $m$ in the set $m \in \{\pm 1, \pm 2\}$ the arguments closely parallel the one just presented and we leave them out of this appendix, preferring to move to the more demanding cases $m<-2$ and $m>2$. 

Referring to Eq. \eqref{edge_on_reflected_phi_integral} we see that the cases $m<-2$ and $m>2$ only differ in their overall sign and the type of trigonometric function acting as a multiplier. We focus on $m<-2$, since the argument for $m>2$ is quite similar. Expanding the sine functions for $m<-2$ we find
	\[
\Phi_{m<-2}(\phi_o) =     
	\frac{-2 \sin(m\phi_o/2)}{m(m^2-4)}\left[\cos\left(\frac{m\pi}{2}\right)\left\{ m\sin \alpha \cos\left(\frac{m\alpha}{2}\right)-2\cos \alpha \sin\left(\frac{m\alpha}{2}\right)\right\} +\sin\left(\frac{m\pi}{2}\right)\left\{m\sin \alpha \sin\left(\frac{m\alpha}{2}\right)+2\cos \alpha \cos\left(\frac{m\alpha}{2}\right)\right\}\right].
	\]
Considering the cases of even $m$ and odd $m$ separately allows us to drop half the terms. For even $m$ this reduces to:
\[
\Phi_{m<-2}(\phi_o) =     
	\frac{2(-1)^{m/2+1} }{m(m^2-4)} \left[ m\sin(\alpha)\cos\left(\frac{m\alpha}{2}\right)\sin\left(\frac{m\phi_o}{2}\right)-2\cos(\alpha)\sin\left(\frac{m\alpha}{2}\right)\sin\left(\frac{m\phi_o}{2}\right) \right],
\]
which is an odd function. For the first term we find:\begin{align*}
b_{n}&=\frac{2(-1)^{m/2+1} }{\pi m(m^2-4)} \int_{-\pi}^{\pi} m\sin(\alpha)\cos\left(\frac{m\alpha}{2}\right)\sin\left(\frac{m\phi_o}{2}\right)\sin\left(n\phi_o\right) d\phi_o \\
&=\frac{2(-1)^{m/2+1} }{\pi m(m^2-4)}\left( \frac{-2mn(1+\cos(n\pi))}{(-1+m-n)(1+m-n)(-1+m+n)(1+m+n)} \right)
\end{align*}
For odd $n$, the numerator always vanishes. As $m<-2$ and $n>1$, the denominator only vanishes for $n=1-m$ and $n=-1-m$; direct integration of these two cases shows that they are proportional to $\sin^2(m \pi)/(m \mp 1)$, respectively, and hence also vanish. 

For the second term, assuming odd $n$ to make trigonometric reductions of the solution faster, we find:\begin{align*}
b_{2n+1}&=\frac{4(-1)^{m/2+1} }{\pi m(m^2-4)} \int_{-\pi}^{\pi} \cos(\alpha)\sin\left(\frac{m\alpha}{2}\right)\sin\left(\frac{m\phi_o}{2}\right)\sin\left((2n+1)\phi_o\right) d\phi_o \\
&=\frac{2(2n+1)^2(1-(2n+1)^2+m^2)\sin\left(\frac{m\pi}{2}\right)^2}{(m - 2 n - 2) (m + 2 n + 2) (m - 2 n) (m + 2 n)}
\end{align*}
Note that as $m$ is even, $\frac{m}{2}$ will always be an integer, and $\sin\left(\frac{m\pi}{2}\right)^2=0$. The denominator is zero only when $n=-\frac{m}{2}$ or $n=-1-\frac{m}{2}$ and direct integration of these two cases yields results proportional to $\sin(m\pi/2)$, which again vanish.

The same strategies can be used for the odd $m$ cases for both $m>2$ and $m<-2$ of term-by-term integration for odd harmonic coefficients. This completes the demonstration that edge-on reflected harmonic light curves do not contain odd Fourier components beyond the fundamental mode. 

% Don't change these lines
\bsp	% typesetting comment
\label{lastpage}
\end{document}